\documentclass[12pt]{article}

\usepackage{amssymb}
\usepackage{amsmath}

\usepackage{latexsym}
\usepackage{graphicx}

\usepackage{cite}

\newcommand{\be}{\begin{equation}}
\newcommand{\ee}{\end{equation}}

\newcommand{\bt}{\beta}

\newcommand{\al}{\alpha}
\newcommand{\ra}{\rightarrow}

\newcommand{\gm}{\gamma}

\newcommand{\cH}{{\cal H}}

\newcommand{\cA}{{\cal A}}
\newcommand{\cB}{{\cal B}}

\newcommand{\rgl}{\rangle}
\newcommand{\lgl}{\langle}

\begin{document}

\begin{center}
{\Large{\bf Positive operator-valued measures\\ in quantum decision theory
} \\ [5mm]

Vyacheslav I. Yukalov$^{1,2}$ and Didier Sornette$^{1,3}$} \\ [3mm]

{\it
$^1$Department of Management, Technology and Economics, \\
ETH Z\"urich, Swiss Federal Institute of Technology, \\
Z\"urich CH-8092, Switzerland \\ 
e-mail: syukalov@ethz.ch \\ [3mm]

$^2$Bogolubov Laboratory of Theoretical Physics, \\
Joint Institute for Nuclear Research, Dubna 141980, Russia \\
e-mail: yukalov@theor.jinr.ru \\ [3mm]

$^3$Swiss Finance Institute, c/o University of Geneva, \\
40 blvd. Du Pont d'Arve, CH 1211 Geneva 4, Switzerland \\
e-mail: dsornette@ethz.ch}

\end{center}

\vskip 2cm

\begin{abstract}
We show that the correct mathematical foundation of quantum decision theory,
dealing with uncertain events, requires the use of positive operator-valued 
measure that is a generalization of the projection-valued measure. The latter
is appropriate for operationally testable events, while the former is 
necessary for characterizing operationally uncertain events. In decision 
making, one has to distinguish composite non-entangled events from composite
entangled events. The mathematical definition of entangled prospects is based
on the theory of Hilbert-Schmidt spaces and is analogous to the definition 
of entangled statistical operators in quantum information theory. We
demonstrate that the necessary condition for the appearance of an interference 
term in the quantum probability is the occurrence of entangled prospects and 
the existence of an entangled strategic state of a decision maker. The origin 
of uncertainties in standard lotteries is explained.   
\end{abstract}

\vskip 5mm


\vskip 5mm

\newpage

\section{Introduction}

Techniques of quantum theory are nowadays widely employed not only for physics 
problems but also in such fields as quantum information processing and quantum
computing [1-5]. Another example is the theory of quantum games [6]. A scheme
of artificial quantum intelligence was suggested [7,8]. Applications of quantum 
techniques to cognitive sciences are also quickly growing. 

Actually, the idea that human decision making could be characterized by quantum
techniques was advanced long ago by Bohr [9,10]. Since then, a number of 
publications have discussed the possibility of using quantum techniques in 
cognitive sciences, as is summarized in the recent books [11-13].

Von Neumann [14] mentioned that the theory of quantum measurements can be 
interpreted as decision theory. There is, however, an important difference
between the standard situation in quantum measurements and the often occurring 
case in realistic decision making. Usual measurements in physics problems are 
operationally testable, resulting in well defined numerical values of the 
measured observables. In decision making, however, it is common to deal with 
composite events requiring to take decisions in uncertain situations. While
the final decision should also be operationally testable, intermediate steps 
often involve uncertainty that is not operationally testable. The correct 
description of such uncertain composite prospects requires the use of more
elaborate mathematics than the projection-valued measure commonly employed
for standard physics problems.      
  
In our previous papers [15-19], we have developed Quantum Decision 
Theory (QDT) whose mathematical basis rests on the theory of quantum 
measurements and quantum information theory. The strategic state of a decision 
maker was represented by a wave function. As is well known, a wave function 
characterizes an isolated quantum system. However, in real life, no decision 
maker can be absolutely isolated from the society where he/she lives. That is, 
the characterization of a strategic decision-maker state by a wave function 
is oversimplified. Simple methods of quantum mechanics are not sufficient for 
a realistic description of a decision maker that is a member of a society. 
Being an open system, a decision maker has to be described by a statistical 
operator, similarly to any non-isolated system in quantum theory, which requires 
the use of the ramified techniques of quantum statistical theory [20-22]. 

One could think that, for describing simple psychological laboratory experiments,
there is no need of invoking statistical operators and it would be sufficient
to just use pure states characterized by wave functions, since most lab-based 
tests on cognition deal with subjects that are typically isolated 
and the experiments have limited durations. However, in 
quantum theory, as is well known, the necessity of using density matrices is 
dictated by the existence of interactions not merely at the present moment of 
time, but also at any previous times, which is always relevant for any alive being. 
In addition, a system is termed open if interactions have been present not 
only with similar systems, but with any surrounding. Thus, humans are always 
subject to interactions with many other people as well as with various information 
sources, such as TV, radio, telephone, internet, newspapers, and so on. All these 
numerous interactions always influence decision makers making them, without doubt, 
open systems. In order to reduce a decision maker to a pure quantum-mechanical 
system for a pure laboratory experiment, it would be necessary to make a surgical 
operation deleting all memory and information from the brains of these poor
decision makers. The usual lab tests, fortunately, do not require this, so that
decision makers have to be always treated as open systems described by statistical 
operators. 

It is worth stressing that a theory based on statistical operators includes, as a
particular case, the pure-state description. So that all results obtained in the 
general consideration can be straightforwardly reduced to the latter case by taking 
the statistical operator in the pure form $|\psi><\psi|$.    
 
The goal of the present report is threefold. First, we present the extension 
of QDT to the most general case when the decision-maker strategic state is 
characterized by a statistical operator. Second, we demonstrate that the correct 
mathematical description of uncertain composite events has to be described by a 
positive operator-valued measure generalizing the projection-valued measure used 
in quantum mechanics. Third, we show that decision making under uncertainty is a 
rather common phenomenon that happens in the delusively simple problem of choosing 
between lotteries and illustrate this by an explicit example. 

The important feature of our approach is that we employ the rigorous mathematical 
techniques developed in the theory of quantum measurements and quantum information
theory. In order that the reader with background in psychology would not confuse
mathematical notions used in our paper, we give the necessary definitions, trying 
at the same time to keep the exposition concise. Details can be found in the 
books [1,2,4,5,23-26 ] and review articles [3,27].

\section{Operationally testable events}

The operationally testable events in QDT can be characterized by analogy 
with operationally testable measurements in quantum theory. Quantum events, 
obeying the Birkhoff-von Neumann quantum logic [28], form a non-commutative
non-distributive ring $\mathcal{R}$. The nonempty collection of all subsets
of the event ring $\mathcal{R}$, including $\mathcal{R}$, which is closed
with respect to countable unions and complements, is the event sigma algebra
$\Sigma$. The algebra of quantum events is the pair $\{\Sigma, \mathcal{R}\}$
of the sigma algebra $\Sigma$ over the event ring $\mathcal{R}$. An elementary
event $A_n$ is represented by a basic state $|n \rgl$ generating the event 
operator defined as a projector 
\be
\label{1}
\hat P_n \equiv   | n \rgl \lgl n | \;  .
\ee
The space of mind of a decision maker is a Hilbert space ${\mathcal H}$ that 
is a closed linear envelope of the basis $\{|n\rgl\}$ spanning all admissible 
basic states. The strategic state of a social decision maker is a statistical 
operator $\hat{\rho}$ that is a trace-class positive operator normalized 
to one. The algebra of observables in QDT is the family $\{\hat{P}_n\}$ of 
the event operators whose expected values define the event probabilities     
\be
\label{2}
p(A_n) \equiv {\rm Tr}\hat\rho \hat P_n \equiv
\lgl \hat P_n \rgl \; ,
\ee
with the trace over $\mathcal{H}$. From this definition, it follows:
$$
 \sum_n p(A_n) = 1 \; , \qquad 0 \leq p(A_n) \leq 1 \;  ,
$$
hence the family $\{ p(A_n) \}$ forms a probability measure. By the Gleason 
theorem [29], this measure is unique for a Hilbert space of dimensionality 
larger than two. In the theory of quantum measurements, the projectors 
$\hat{P}_n$ play the role of observables, so that, for an event $A_n$, one 
has the correspondence
\be
\label{3}
 A_n \ra \hat P_n \equiv | n \rgl \lgl n | \;  .
\ee
For a union of mutually orthogonal events, there is the correspondence
\be
\label{4}
 \bigcup_n A_n ~ \ra ~ \sum_n \hat P_n \;  ,
\ee
which results in the additivity of the probabilities:
\be
\label{5}
 p\left ( \bigcup_n A_n \right ) = \sum_n p(A_n) \;  .
\ee
The procedure described above is called the standard measurement.

Let us emphasize that all formulas of this and following sections
are valid for an arbitrary statistical operator, which includes the pure 
form $\hat\rho = \vert \psi \rgl \lgl \psi \vert$ as a trivial particular 
case.

\section{Operationally uncertain events}

It may happen that one cannot tell whether a particular event has occurred, 
but it is only known that some of the events $A_n$ could be realized. This 
is what is called an uncertain or inconclusive event.

Assume that the observed event $A$ is a set $\{A_n\}$ of possible events. 
Although the events $A_m$ and $A_n$ are orthogonal for $m \neq n$, in 
the case of the uncertain event $A$, they form not a standard union but an 
{\it uncertain union} that we shall denote as
\be
\label{6}
 A \equiv \{ A_n \} \equiv \biguplus_n A_n
\ee
in order to distinguish it from the standard union $\bigcup_n A_n$. The
uncertain event $A$ is characterized by the wave function
\be
\label{7}
 | A \rgl = \sum_n a_n | n \rgl \;   ,
\ee
where $|a_n|^2$ play the role of weights for the events $A_n$. Now, instead 
of the correspondence (4) for the standard union, we have the correspondence
\be
\label{8}
 \biguplus_n A_n ~ \ra ~ \hat P_A \equiv | A \rgl \lgl A | \;  .
\ee
Note that $\hat P_A$ is not a projector.

The probability of the uncertain event $A$ reads as
\be
\label{9}
 p(A) = p\left ( \biguplus_n A_n \right ) = \sum_n | a_n |^2 p(A_n)
+ q(A) \;  ,
\ee
where the second term
\be
\label{10}
q(A) \equiv \sum_{m\neq n} a_m^* a_n \lgl m | \hat\rho | n \rgl
\ee
is caused by the interference of the uncertain subevents $A_n$ that are
called {\it modes}. The probability $p(A)$ of the uncertain event $A$, 
represented by the uncertain union (6), does not equal the sum of the
event probabilities $p(A_n)$. In that sense, the uncertain union (6) is  
not additive with respect to partial events $A_n$, contrary to the 
probability of the standard union (5).

\section{Composite non-entangled prospects}

Composite events are termed prospects. These can be sorted
in two classes, entangled 
and non-entangled [30]. This classification is based on the theory of 
Hilbert-Schimdt spaces [31,32], as is explained below. 

It is well known that quantum-mechanical wave functions, pertaining to a Hilbert 
space, can be either entangled, or non-separable, and non-entangled, or separable.
Similarly, by constructing the appropriate Hilbert-Schmidt space, it is possible 
to introduce the notions of entangled, or non-separable operators, and of 
non-entangled, or separable operators. We need this classification for a bipartite 
system, that is, consisting of two parts, although the definition can be 
straightforwardly generalized for a multi-partite case.

Let us consider a bipartite quantum system, with one subsystem corresponding to
a Hilbert space $\mathcal{H}_A$ and the other, to a Hilbert space $\mathcal{H}_B$.
The subsystem, defined in $\mathcal{H}_A$, is characterized be a set of operators, 
acting on $\mathcal{H}_A$. For what follows, we keep in mind the operators of 
observables represented by the projectors of operationally testable events. The 
set of operators on $\mathcal{H}_A$ forms an algebra of observables $\mathcal{A}$. 
Respectively, the subsystem in $\mathcal{H}_B$ is characterized by an algebra 
$\mathcal{B}$ acting on $\mathcal{H}_B$. For any two operators $\hat{A}_1$ and 
$\hat{A}_2$ in the algebra $\mathcal{A}$, it is possible to introduce a scalar 
product  
$$
(\hat A_1 , \hat A_2 ) \equiv {\rm Tr}_A \hat A_1^+ \hat A_2 \; ,
$$
where the trace is over the space $\mathcal{H}_A$, inducing the Hilbert-Schimdt 
norm
$$
|| \hat A|| \equiv \sqrt{ ( \hat A, \hat A ) } \;   .
$$
Then the operator algebra $\mathcal{A}$, complemented by the above scalar product, 
becomes a Hilbert-Schmidt space. The same can be done for the algebra $\mathcal{B}$
becoming a Hilbert-Schmidt space with the scalar product defined in the same way.

The system, composed of two parts, is a composite system defined in the tensor-product
Hilbert space $\cH_A \bigotimes \cH_B$. The operator algebra $\cA \bigotimes \cB$
acts on this tensor-product space. For any two operators $\hat{C}_1$ and $\hat{C}_2$
of the latter algebra, one defines the scalar product
$$
(\hat C_1 , \hat C_2 ) \equiv {\rm Tr}_{AB} \hat C_1^+ \hat C_2 \;   ,
$$
with the trace over the space $\cH_A \bigotimes \cH_B$. Thus, the algebra 
$\cA \bigotimes \cB$, complemented by this scalar product, becomes a composite 
Hilbert-Schmidt space. In this way, the operators of a Hilbert-Schmidt space can 
be treated similarly to the vectors of a Hilbert space. 

One tells that the operator $\hat{C}$, acting on the composite Hilbert-Schmidt 
space, is separable, or not-entangled, if and only if it can be represented as 
$$
\hat C = \sum_\gm \hat C_{\gm A} \bigotimes \hat C_{\gm B} \;   ,
$$
where $\hat{C}_{\gamma A}$ and $\hat{C}_{\gamma B}$ are the operators from the 
related algebras of observables, acting on $\cH_A$ and $\cH_B$, respectively. Such 
separable operators have been widely used in scattering theory [33]. On the contrary, 
if the operator cannot be reduced to the separable form, it is termed non-separable, 
or entangled.

The classification of the operators onto separable and entangled is intensively 
employed in quantum information theory [1-5], where one considers statistical 
operators. In quantum decision theory [15-19], this classification is applied to 
prospect operators. A prospect operator that cannot be represented in the separable 
form is called entangled or non-separable, while when it can be reduced to that form, 
it is termed non-entangled, or separable. Similarly, the prospects, represented by 
the corresponding prospect operators, can be distinguished onto entangled and 
non-entangled. Exactly this classification will be used below.              
     
Suppose we consider two elementary events $A_n$, represented by a vector 
$| n \rgl$ from a Hilbert space $\mathcal{H}_A$, and $B_\alpha$, represented 
by a vector $| \al \rgl$ from a Hilbert space $\mathcal{H}_B$. The composite 
event, formed by these two elementary events, is treated as a tensor product 
$A_n \bigotimes B_\alpha$. In this notation, the event $A_n$ is assumed to 
happen after the event $B_\alpha$. The composite event $A_n \bigotimes B_\alpha$
is represented by the vector 
\be
\label{11}
 | n \alpha \rgl = | n \rgl \bigotimes | \al \rgl
\ee
from the tensor-product Hilbert space 
\be
\label{12}
 \cH_{AB} \equiv \cH_A \bigotimes \cH_B \;  .
\ee
The composite event of observing $A_n$ and $B_\alpha$ induces the correspondence
\be
\label{13}
 A_n \bigotimes B_\al ~ \ra ~ \hat P_n \bigotimes \hat P_\al \; ,
\ee
where
$$
 \hat P_\al \equiv | \al \rgl \lgl \al |
$$
is a projector in $\mathcal{H}_B$.

The strategic state is now a statistical operator $\hat{\rho}$ on the
tensor-product space (12). The joint probability of the composite event (13)
becomes
\be
\label{14}
  p(A_n \bigotimes B_\al ) = {\rm Tr} \hat \rho \hat P_n \bigotimes
\hat P_\al \equiv \lgl \hat P_n \bigotimes \hat P_\al \rgl \;  ,
\ee
with the trace over the space (12). The composite event (13) is the simplest
composite event, which enjoys the factor form, being composed of two
elementary events, and being called {\it non-entangled}.  

More complicated structures arise when at least one of the events is a union. 
It is important to emphasize the difference between the standard union and 
the uncertain union introduced in (6).

When the composite event is a product of an elementary event $A_n$ and a 
standard union of mutually orthogonal events $B_\alpha$, we can employ the
known property of composite events:
$$
 A_n \bigotimes \; \bigcup_\al B_\al =
\bigcup_\al A_n \bigotimes B_\al \; .
$$
In the right-hand side here, we have the union of mutually orthogonal composite 
events, since $B_\alpha$ are assumed to be mutually orthogonal. Therefore
\be
\label{15}
 p \left ( A_n \bigotimes \bigcup_\al B_\al \right ) =
\sum_\al p(A_n \bigotimes B_\al ) \; .
\ee
That is, the probability of a composite event, with one of the factors being
the standard union of mutually orthogonal events, is additive. Such events are
also termed non-entangled. 

It is important to stress that the used terminology is in agreement with the 
notions of separable and non-entangled operators, as is formulated at the beginning 
of this section. Really, in the present case, the bipartite system is described 
by the tensor product of the algebras of observables 
$\mathcal{A} \equiv \{ \hat{P}_n \}$ and $\mathcal{B} \equiv \{ \hat{P}_\alpha \}$. 
The event $A_n \bigotimes \; \bigcup_\al B_\al$ induces the event operator that
has the form
$$
\hat P \left ( A_n \bigotimes \bigcup_\al B_\al \right ) =
\sum_\al \hat P_n \bigotimes \hat P_\al \; ,
$$
corresponding to the definition of a separable, or non-entangled operator.

\section{Composite entangled prospects}

The situation is essentially different when dealing with an uncertain union. 
In that case, composite events are represented by prospect operators that
cannot be represented in the separable form.

Let us have such an uncertain union
\be
\label{16}
 B \equiv \{ B_\al \} \equiv \biguplus_\al B_\al
\ee
corresponding to a vector
\be
\label{17}
 | B \rgl = \sum_\al b_\al | \al \rgl \; .
\ee
The composite event, or prospect
\be
\label{18}
 \pi_n =  A_n \bigotimes B = A_n \bigotimes \biguplus_\al B_\al \; ,
\ee
corresponds to the prospect state
\be
\label{19}
  | \pi_n \rgl = | n \rgl \bigotimes | B \rgl =
\sum_\al b_\al | n \al \rgl \; .
\ee
The composite event (18) induces the correspondence
\be
\label{20}
\pi_n ~ \ra ~ \hat P(\pi_n) \equiv | \pi_n \rgl \lgl \pi_n |
\ee
defining the prospect operator $\hat{P}(\pi_n)$. 

At this point, it is necessary to make an important comment clarifying
the notion of entanglement. The latter is correctly defined when it is
explicitly stated what object is considered and with respect to which
parts it is entangled or not. Recall that, in quantum mechanics, one 
considers the entanglement of a wave function of a composite system
with respect to the wave functions of its parts, which are not arbitrary 
functions. Wave functions can be defined as eigenfunctions of Hamiltonians. 
In the case of statistical operators, one considers their entanglement 
or separability with respect to statistical operators of subsystems, 
but not with respect to arbitrary operators [1-5]. In the general case 
of operators from a Hilbert-Schmidt space, one considers their 
entanglement with respect to the operators from the corresponding 
Hilbert-Schmidt subspaces, but not with respect to arbitrary operators
not pertaining to the prescribed subspaces. 

The prospect state (19) could be qualified as not entangled with respect 
to arbitrary functions from the Hilbert spaces $\mathcal{H}_A$ and
$\mathcal{H}_B$. However, the function $\vert B \rgl$ does not correspond
to an operationally testable event, while exactly the latter are of our 
interest. Hence the formal separability of (19) in the Hilbert space is
of no importance. What is important is the separability or entanglement 
of operationally testable events. 

In our case, the operationally testable events correspond to the related 
projectors forming the algebras of observables 
$\mathcal{A} \equiv \{ \hat{P}_n \}$ and $\mathcal{B} \equiv \{ \hat{P}_\alpha \}$.
Complementing them by the appropriate scalar products, we get the 
corresponding Hilbert-Schmidt spaces. A prospect $\pi_n$ is characterized by 
the prospect operator (20). According to the general theory, a prospect 
operator is separable if and only if it can be reduced to the linear 
combination of the tensor products of operators from the Hilbert-Schmidt
subspaces. However, the prospect operator (20) reads as
$$
\hat P(\pi_n) = \sum_\al \; | b_\al|^2 \hat P_n \bigotimes \hat P_\al \; 
+ \; \sum_{\al\neq \bt} 
b_\al b_\bt^* \hat P_n \bigotimes | \al \rgl \lgl \bt | \;  .
$$
The first term in the right-hand side does correspond to the definition of
separability, while the second term does not, since 
$\vert \alpha \rgl \lgl \beta \vert$, with $\alpha \neq \beta$, does not pertain 
to the algebra of observables $\mathcal{B} \equiv \{ \hat{P}_\alpha \}$.  
Hence, the prospect operator (20) is not separable, that is, it is entangled. 
Respectively, prospect (18), corresponding to this prospect operator can also 
be termed entangled, since it cannot be represented as a union of mutually 
orthogonal events. The entangling properties of operators can be quantified 
by the measure of entanglement production [34-36]. The amount of entanglement
produced in the process of decision making can be calculated as shown in Ref. [37].   

The prospect states (19) are not necessarily orthogonal and do not need
to be normalized to one. Because of this, the prospect operators, generally, 
are not projectors. However, the resolution of unity is required:
\be
\label{21}
 \sum_n \hat P(\pi_n) = \hat 1_{AB} \;  ,
\ee
where $\hat{1}_{AB}$ is the unity operator in the space (12). The family
$\{\hat{P}(\pi_n)\}$ composes a {\it positive operator-valued measure}.

The prospect probability
\be
\label{22}
p(\pi_n) \equiv {\rm Tr} \hat\rho \hat P(\pi_n) \equiv
\lgl \hat P(\pi_n) \rgl \; ,
\ee
with the trace over space (12), becomes the sum of two terms:
\be
\label{23}
 p(\pi_n) = f(\pi_n) + q(\pi_n) \;  .
\ee
The first term
\be
\label{24}
 f(\pi_n) \equiv \sum_\al | b_\al |^2 p(A_n \bigotimes B_\al ) \; 
\ee
contains the diagonal elements with respect to $\alpha$, while the second term
\be
\label{25}
 q(\pi_n) \equiv \sum_{\al\neq\bt} b_\al^* b_\bt
\lgl n\al | \hat\rho | n \bt \rgl \;  
\ee
is formed by the nondiagonal elements. By construction and due to the resolution
of unity (21), the prospect probability (22) satisfies the properties
\be
\label{26}
 \sum_n p(\pi_n) = 1 \; , \qquad 0 \leq p(\pi_n) \leq 1 \;  ,
\ee
which makes the family $\{p(\pi_n)\}$ a probability measure.

Expression (25) is caused by the quantum nature of the considered events
producing interference of the modes composing the uncertain union (16).
Because of this, the term (25) can be called the {\it quantum factor}, 
{\it interference factor}, or {\it coherence factor}. The quantum term (25)
may be nonzero only when prospect (18) is {\it entangled} in the sense of
the nonseparability of the prospect operator in the Hilbert-Schmidt space.  

Classical probability has to be a marginal case of quantum probability. 
To this end, we have to remember the quantum-classical correspondence 
principle advanced by Bohr [38]. This principle tells us that classical theory 
is to be the limiting case of quantum theory, when quantum effects vanish. 
In the present case, this implies that when the quantum interference factor 
tends to zero, the quantum probability has to tend to a classical probability. 
Such a process is also called {\it decoherence}. According to the principle
of the quantum-classical correspondence, we have
\be
\label{27}
 p(\pi_n) ~ \ra ~ f(\pi_n) \; , \qquad q(\pi_n) ~ \ra ~ 0 \;  ,
\ee
which means that the decoherence process leads to the classical probability
$f(\pi_n)$. Being a probability, this classical factor needs to be normalized,
so as to satisfy the conditions
\be
\label{28}
 \sum_n f(\pi_n) = 1 \; , \qquad 0 \leq f(\pi_n) \leq 1 \;  .
\ee
As a consequence of the above equations, the interference factor enjoys the
properties
\be
\label{29}
 \sum_n q(\pi_n) = 0 \; , \qquad - 1 \leq q(\pi_n) \leq 1 \;  .
\ee
The first of these equations is called the {\it alternation condition}.

One should not confuse the effect of decoherence, based on the quantum-classical
correspondence principle, when quantum measurements are reduced to classical, 
with the Kochen-Specker theorem [39] stating the impossibility of simultaneous 
embedding of all commuting sub-algebras of the algebra of quantum observables 
in one commutative algebra, assumed to represent the classical structure of the 
hidden-variables theory, if the Hilbert space dimension is at least three.
This theorem places certain constraints on the permissible types of hidden-variable 
theories, which try to explain the apparent randomness of quantum mechanics as a 
deterministic model featuring hidden states. The theorem excludes hidden-variable 
theories that require elements of physical reality to be non-contextual, i.e., 
independent of the measurement arrangement. The exclusion of such hidden variables 
is exactly due to the existence of quantum entanglement. 

Our consideration has nothing to do with hidden variables. We do not intend to 
replace quantum theory by a classical theory with hidden variables. Vice versa, 
the whole of our approach is completely based on the standard techniques of quantum 
theory and all results are in full agreement with the known properties of quantum 
theory. Being always in the frame of quantum theory, we consider a very well known 
effect called {\it decoherence} that manifests the transition from quantum to classical 
behavior. This effect is intimately connected with the quantum-classical correspondence 
principle, formulated by Bohr and widely used in quantum theory. According to this 
principle, the results of quantum theory reduce to those of classical theory, when 
quantum effects, such as entanglement and interference, are washed out. The effect
of decoherence is well understood and described in the frame of quantum theory [40-42].  

As is mentioned above, the quantum term arises only when the considered prospect 
is entangled in the sense of the nonseparability of the prospect operator in the 
Hilbert-Schmidt space. The other necessary condition for the existence of the quantum 
term is the entanglement in the strategic state $\hat{\rho}$. To illustrate that a 
disentangled strategic state does not produce interference, let us take the system 
state in the disentangled product form
$$
 \hat \rho = \hat \rho_A \bigotimes \hat \rho_B \;  .
$$
Then, the quantum interference term becomes
$$
q(\pi_n) = \sum_{\al\neq\bt} b_\al^* b_\bt \lgl n | \hat\rho_A | n \rgl
\lgl \al | \hat \rho_B | \bt \rgl \;   .
$$
Taking into account the normalization condition 
$$
{\rm Tr}_A \hat\rho_A = \sum_n  \lgl n | \hat\rho_A | n \rgl = 1 \;  ,
$$
we get
$$
 \sum_n q(\pi_n) =
\sum_{\al\neq\bt} b_\al^* b_\bt \lgl \al | \hat\rho_B | \bt \rgl = 0 \; .
$$
As a result, we find
$$
 q(\pi_n) = \lgl n | \hat\rho_A | n \rgl \sum_n q(\pi_n) = 0 \;  .
$$
So, the disentangled strategic state does not allow for the appearance of a 
nontrivial quantum interference term.    

The quantum term (25) is a random quantity satisfying a very important 
property called the {\it quarter law} [16-19]. For a prospect lattice,
${\mathcal L} \equiv \{ \pi_n: n = 1, 2, \ldots N_L \}$ the absolute value 
of the aggregate quantum factor can be estimated as
\be
\label{30}
|\bar{q}| \equiv \frac{1}{N_L} \; \sum_{j=1}^{N_L} \; |\; q(\pi_j) \; | = 
\frac{1}{4} \; .
\ee
The value $1/4$ for the aggregate attraction factor can be shown [43] to 
arise for a large class of distributions characterizing the attraction factors 
of different decision makers. 

Expression (24), corresponding to classical probability, is defined as an 
objective term, whose value is prescribed by the prospect utility, 
justifying to call this term the {\it utility factor}. The quantum term (25)
describes the attractiveness of the prospect to a decision maker, so that it
is named the {\it attraction factor} [15-19]. A prospect $\pi_1$
is more useful than $\pi_2$, if and only if $f(\pi_1) > f(\pi_2)$. A prospect 
$\pi_1$ is more attractive than $\pi_2$, when and only when $q(\pi_1) > q(\pi_2)$.
And a prospect $\pi_1$ is preferable to $\pi_2$, if and only if 
$p(\pi_1) > p(\pi_2)$. Hence, a prospect can be more useful but less attractive, 
as a result being less preferable.

\section{Uncertainty in standard lotteries}

It can be shown [30], that the necessary condition for the quantum term to be
nonzero requires that the considered prospect be entangled and the strategic
state $\hat{\rho}$ also be entangled. This implies that the decision is made
under uncertainty [44]. 

A typical situation in decision making is when one chooses between several 
lotteries. One may ask what kind of uncertainty is ascribed to such a choice 
between the lotteries. 

Suppose we consider a family $\{L_n\}$ of lotteries enumerated with the index 
$n = 1, 2, \ldots, N_L$. Each lottery is the set
\be
\label{31}    
L_n = \{ x_i , \; p_n(x_i): \; i=1,2,\ldots \}
\ee
of payoffs $x_i$ and payoff probabilities $p_n(x_i)$. A decision maker has to
choose one of these lotteries. 

The choice between the lotteries is a random procedure involving uncertainty. 
First of all, when choosing a lottery, one does not know exactly what would 
be a payoff whose occurrence is characterized by the related probability. 
Moreover, in each choice, there always exists uncertainty caused by two reasons.
One reason is the decision-maker doubt about the objectivity of the setup 
suggesting the choice. The other origin of uncertainty is caused by  
subjective hesitations of the decision maker with respect to his/her correct 
understanding of the problem and his/her knowledge of what would be the best 
criterion for making a particular choice. Let us denote by $B_1$ the 
decision-maker confidence in the empirical setup as well as in his/her ability
of making a correct decision. Then $B_2$ corresponds to the disbelief of the 
decision maker in the suggested setup and/or in his/her understanding of the 
appropriate criteria for the choice. The combination of belief and disbelief 
is the set
$$
B = \{ B_1 , B_2 \} = \biguplus_\al B_\al \qquad ( \al=1,2 ) \;  .
$$  

In this way, even choosing between simple lotteries $L_n$, one actually 
confronts the composite prospects
\be
\label{32}
 \pi_n = L_n \bigotimes B \;  ,
\ee
where the event of selecting a lottery $L_n$ is denoted by the same letter,
which should not lead to confusion. The choice is made under uncertainty
incorporated into the set $B = \{B_1, B_2\}$ of belief and disbelief. 

The prospect probability is given by (23). The utility factor, characterizing 
the objective part of the probability, in the case of the choice between the 
lotteries can be defined [16,19] as
\be
\label{33}
 f(\pi_n) = \frac{U(\pi_n)}{\sum_n U(\pi_n) } \;  ,
\ee
ordering the prospects according to their expected utilities
\be
\label{34}
 U(\pi_n) = \sum_i u(x_i) p_n(x_i) \;  ,
\ee
where $u(x_i)$ is a utility function. The attraction factors can be evaluated
as is explained in the previous section. 

For example, dealing with the prospect lattice $\mathcal{L} = \{ \pi_1, \pi_2 \}$,
in which the prospect $\pi_i$ is more attractive than $\pi_j$, the prospect 
probabilities are estimated by the expressions
$$
p(\pi_i) = f(\pi_i) + 0.25 \; ,
$$
\be
\label{35}
 p(\pi_j) = f(\pi_j) - 0.25 \;  .
\ee
 
To illustrate how the procedure described above works, let us consider the
lotteries discussed by Kahneman and Tversky [45]. Consider, for instance,
the lotteries
$$
 L_1 = \{ 6, 0.45\; | \; 0, 0.55  \} \; , \qquad
 L_2 = \{ 3, 0.9 \; | \; 0, 0.1 \} \;   .
$$
Calculating their expected utilities, we assume, for simplicity, linear
utility functions $u(x) = c x$. The corresponding utility factors of both 
the lotteries are equal,
$$
 f(\pi_1) = 0.5 , \qquad f(\pi_2) = 0.5 \;  .
$$
The second prospect is more attractive, being more certain. Then, employing 
rule (35), we have
$$
  p(\pi_1) = 0.25 , \qquad p(\pi_2) = 0.75 \;  .
$$
The experimental results of Kahneman and Tversky [45] are
$$
p_{exp}(\pi_1) = 0.14 , \qquad p_{exp}(\pi_2) = 0.86 \; ,
$$
where $p_{exp}(\pi_i)$ is the ratio of the number of the decision makers, 
choosing the lottery $L_i$, to the total number of participants. Within the 
statistical errors of $\pm 0.1$ of these experiments, our theoretical 
prediction agrees with the empirical results.

Another example by Kahneman and Tversky [45] is the choice between the lotteries
$$
L_1 = \{ 6, 0.001\; | \; 0, 0.999  \} \; , \qquad
L_2 = \{ 3, 0.002 \; | \; 0, 0.998 \} \; ,
$$
which enjoy the same utility factors
$$
f(\pi_1) = 0.5 , \qquad f(\pi_2) = 0.5 \;   ,
$$
as in the previous example. The uncertainties of the two lotteries are close
to each other. However, the gain in the first prospect is essentially
larger, which makes it more attractive, hence, the second prospect less attractive. 
As a result, the prospect preference reverses, as compared to the previous case, 
with the prospect probabilities
$$
 p(\pi_1) = 0.75 , \qquad p(\pi_2) = 0.25 \;   .
$$
The experimental data of Kahneman and Tversky [45] are
$$
p_{exp}(\pi_1) = 0.73 , \qquad p_{exp}(\pi_2) = 0.27 \; .
$$
Thus our theoretical prediction practically coincides with the empirical data.

We also have analyzed a number of other experimental examples, obtaining  
good agreement of our theoretical predictions with empirical results. However, 
we shall not overload the present report by the description of all these 
experiments, which will be the topic of a separate paper.

\section{Conclusion}

Decision making very often meets the necessity of deciding under uncertainty.
Applying quantum techniques to decision making, one has to use the appropriate
mathematical tools. We have shown that the correct mathematical foundation 
of quantum decision theory, dealing with uncertain events, requires the use of 
positive operator-valued measure that is a generalization of the projection-valued 
measure. The latter is used for operationally testable events, but cannot be 
applied to uncertain events typical of decision making under uncertainty. Such 
operationally uncertain events require the use of the operator-valued measure. 
In decision making, one has to distinguish composite non-entangled events from 
composite entangled events. The accurate mathematical formulation of entangled 
events is based on the notion of entangled or nonseparable prospect operators 
in a Hilbert-Schmidt space. This should not be confused with the entanglement
of functions in a Hilbert space. The operationally testable events are called 
modes. Therefore the entanglement of such events can be termed 
{\it mode entanglement}. 

According to the principle of quantum-classical correspondence, classical 
probabilities can be treated as a limiting case of quantum probabilities, when
the effect of decoherence is present. We consider the occurrence of classical
probabilities exactly in this sense, which is completely in the frame of quantum
theory. We stress that the Kochen-Specker theorem, proving the impossibility of
non-contextual hidden variables has no relation to our approach.
    
Quantum probability can be essentially different from the form of classical 
probability only for entangled events, defined through the mode entanglement in 
a Hilbert-Schmidt space. The necessary condition for the appearance of a quantum 
interference term in the quantum probability is the occurrence of entangled 
prospects and the existence of an entangled strategic state of a decision maker. 
The origin of uncertainties in standard lotteries is explained. Our approach makes 
it possible to provide theoretical predictions that are in good numerical 
agreement with the results of empirical observations.      

Our approach is principally different from those of other authors [11] in several 
basic points. First, we develop a general theory based on rigorous mathematics
of quantum measurement theory and quantum information theory, which can be applied 
to any decision-making processes. Different from [11] and others, we do not construct special schemes for studying some particular problems. Second, we give a mathematically correct definition of quantum joint probabilities as the probabilities of composite
events realized in different measurement channels and represented in tensor-product 
Hilbert spaces. In contrast, other authors usually consider a single Hilbert space and 
deal with the L\"{u}ders probabilities of consecutive events, which are, actually, 
transition probabilities, but cannot be treated as conditional probabilities [30].    
Third, we emphasize the necessity of entangled events for the appearance of quantum 
effects, such as the arising interference. Fourth, we define the entanglement of
operationally testable events as mode entanglement described by the nonseparable
prospect operators in a Hilbert-Schmidt space. Fifth, our theory allows for
quantitative predictions of decision making, without any fitting parameters, 
including quantitative explanations of classical decision-making paradoxes. This
makes our approach unique, since there is no other approach that could 
be compared with empirical results without invoking fitting parameters.

\end{document}